# Image detection–based high-throughput sorting of particles using traveling surface acoustic waves in microscale flows

Nikhil Sethia[1], Joseph Sushil Rao[2,3], Amit Manicka[4,5], Michael L. Etheridge[4], Erik B. Finger[2], John C. Bischof[4,6], Cari S. Dutcher*[1,4]

[1]Department of Chemical Engineering and Materials Science, University of Minnesota, Minneapolis, MN, USA

[2]Division of Solid Organ Transplantation, Department of Surgery, University of Minnesota, Minneapolis, MN, USA

[3]Schulze Diabetes Institute, Department of Surgery, University of Minnesota, Minneapolis, MN, USA

[4]Department of Mechanical Engineering, University of Minnesota, Minneapolis, MN, USA

[5]Department of Computer Science and Engineering, University of Minnesota, Minneapolis, MN, USA

[6]Department of Biomedical Engineering, University of Minnesota, Minneapolis, MN, USA

*Corresponding author (email: cdutcher@umn.edu)

**Abstract**

Large particle sorters have potential applications in sorting microplastics and large biomaterials (>50 μm), such as tissues, spheroids, organoids, and embryos. Though great advancements have been made in image-based sorting of cells and particles (<50 μm), their translation for high-throughput sorting of larger biomaterials and particles (>50 µm) has been more limited. An image-based detection technique is highly desirable due to richness of the data (including size, shape, color, morphology, and optical density) that can be extracted from live images of individualized biomaterials or particles. Such a detection technique is label-free and can be integrated with a contact-free actuation mechanism such as one based on traveling surface acoustic waves (TSAWs). Recent advances in using TSAWs for sorting cells and particles (<50 µm) have demonstrated short response times (<1 ms), high biocompatibility, and reduced energy requirements to actuate. Additionally, TSAW-based devices are miniaturized and easier to integrate with an image-based detection technique. In this work, a high-throughput image-detection based large particle microfluidic sorting technique is implemented. The technique is used to separate binary mixtures of small and large polyethylene particles (ranging between ~45–180 µm in size). All particles in flow were first optically interrogated for size, followed by actuations using momentum transfer from TSAW pulses, if they satisfied the size cutoff criterion. The effect of control parameters such as duration and power of TSAW actuation pulse, inlet flow rates, and sample dilution on sorting efficiency and throughput was observed. At the chosen conditions, this sorting technique can sort on average ~4.9–34.3 particles/s (perform ~2–3 actuations/s), depending on the initial sample composition and concentration.

## 1. Introduction

Sorting particles and biomaterials with diameters >50$\mu m$ has applications in environmental remediation, biomedical research, and clinical settings. Research efforts have been made to sort heterogeneous mixtures of microplastics (Akiyama et al. 2020; Perera and Piyasena 2022; Jonai et al. 2023), embryoid bodies (Lillehoj et al. 2010; Buschke et al. 2012; Buschke et al. 2013), drosophila embryos (Furlong et al. 2001; Utharala et al. 2018), zebrafish embryos (Diouf 2025), tumor organoids (Yan et al. 2025), stem cell-derived β cell clusters (SC-β cell clusters) (Sethia et al. 2024), and pancreatic islets of Langerhans (Fernandez et al. 2005; Nam et al. 2010; Steffen et al. 2011). Samples of such particles and biomaterials often have heterogeneity in properties such as size, shape, morphology, genotype, and protein expression, which can be a selective criterion for a given application. Identifying and separating the particles and biomaterials according to such properties calls for individual inspection. An image-based identification technique is highly preferred, as an abundance of information about individual particles and biomaterial can be extracted in a label-free manner. One such detail is the size of the particle/biomaterial, which has been a particularly important parameter in sorting microplastics (up to ~200$\mu$m), embryoid bodies (up to ~400$\mu$m), tumor organoids (up to ~200$\mu$m), SC-β cell clusters (up to ~500$\mu$m) and islets (up to ~350$\mu$m).

While traditional flow cytometers are used for analyzing and sorting cells and small particles (<50$\mu$m), Union Biometrica's COPAS instrument can sort biomaterials of sizes up to 1500$\mu$m with a throughput of 10–30 biomaterials/s (Steffen et al. 2011; Isozaki et al. 2020; LaBelle et al. 2021). Accessibility of COPAS instrument is limited for everyday basic research due to cost and space considerations. Another commercial sorting device, Bionomous's EggSorter (https://bionomous.ch/bioeggsorter/), can sort biological entities of sizes ~500 to 2000$\mu$m, but yields a lower throughput of ~0.25-0.50 entity/s. Image based sorting of large biological systems can also be achieved through use of robotics, however such techniques also lack in throughput (taking >2s per separation event) (Graf 2011; Breitwieser et al. 2018; Diouf et al. 2024; Stepanov et al. 2025). The advancements in microfluidic device fabrication over the past few decades have rapidly miniaturized devices, increasing their accessibility. While the field of sorting of cells and particles (<50$\mu$m) has made progress (Xi et al. 2017), the tools to separate larger particles and biomaterials (>50$\mu$m) have been limited. Manipulating large particles and biomaterials requires a design with larger channels to support higher flow rates wherein fluid inertial effects might become significant. Scaling up the operation is not always intuitive due to the non-linear nature of the Navier–Stokes equations and the corresponding manipulating forces. Additionally, there can be challenges in precisely displacing and collecting the samples due to increased sample mass (Hansen 2002).

Few microfluidic platforms designed for sorting large particles and biosystems (>50$\mu$m) by properties (such as size) lacked optical detection (Lillehoj et al. 2010; Nam et al. 2010; Miller et al. 2016; Gong et al. 2024). Additionally, most of these microfluidic platforms were passive sorting devices, which require laborious and costly redesigning efforts for each sorting criterion. However, with an active sorting technique, there is more control over the sorting process with easily adjustable control parameters to suit the specific needs. Diouf et al. sorted zebrafish embryos into 3 categories through real-time image-based classification and actuation using peristaltic pumps with an average sorting time of 2.92 s per embryo (Diouf 2025). Active sorting techniques can also use software-controlled valves at channel outlets to redirect the sample flow into the desired outlets (Buschke et al. 2012; Buschke et al. 2013; Utharala et al. 2018; Yan et al. 2025). Buschke et al. designed a real-time image segmentation-based particle and embryoid bodies sorting technique (Buschke et al. 2012; Buschke et al. 2013). However, the actuation time of 0.1–1 s, a time delay of ~3 s, and low scan speed of ~5 frames/s limited the throughput of the device. Similarly, Utharala et al. used camera images to sort Drosophila embryos (Utharala et al. 2018). The throughput was limited to 0.13 embryos/s. Yan et al. filtered ~50-150$\mu$m sized organoids from a sample containing ~30-200$\mu$m sized organoids, where organoid sizes were detected from their real-time images. However, the sorting throughput was limited to 0.22 organoids/s due to long valve switching time (~4s) (Yan et al. 2025). Other novel mechanisms to separate large size particles and biological systems using optical detection exist. However, these mechanisms need to be further tested over larger sample size (~order of 10 or higher) to quantify sorting efficiency and throughput (Dang et al. 2024; Wang et al. 2025).

Acoustic-based microfluidic technologies provide excellent biocompatibility, with high durability due to no moving components like a valve, and a particularly low response time compared to other microfluidic methods described above (Xi et al. 2017). Efforts towards sorting cells and small particles (<50$\mu$m) using acoustic controls have been effective, while large-particle image-based sorting is still a developing field. Large-particle actuators are often based on bulk acoustic wave (BAW) transducers (Johnson and Feke 1995; Nilsson et al. 2004; Perera and Piyasena 2022; Jonai et al. 2023). Using BAW-based devices may pose challenges with selective actuation of the desired sample, higher energy requirements, imaging, and miniaturization (Ding et al. 2013). These are due to the wide sorting region (≫1mm), the energy-intensive transmission of the waves through the device, use of opaque substrates, and the large size of the actuator and the device. Additionally, such devices can be challenging to integrate with other microfluidic applications, as they are usually fabricated of glass or silicon (Nilsson et al. 2004).

Alternatively, devices that use surface acoustic waves (SAW) can overcome these limitations. They can be more energy efficient, as wave energy can be confined to the surface of the substrate. They can be easily imaged through, as they are frequently fabricated using transparent materials (such as lithium niobate base and polydimethylsiloxane (PDMS) walls). SAW-based devices can use standing surface acoustic wave (SSAW) (Shi et al. 2009; Nam et al. 2011; Li et al. 2013) or traveling surface acoustic wave (TSAW) (Collins et al. 2016; Ma et al. 2016; Ma et al. 2017; Mutafopulos et al. 2019; Li et al. 2019; Afzal et al. 2020; Li and Ai 2021; Li et al. 2021; Nawaz et al. 2023) for sorting. SSAW-based devices require precise alignment of the bonding surfaces during the fabrication, which is challenging. They are also limited by maximum particle displacement (quarter wavelength of the acoustic field), which can be a challenge for large particles as they require long displacement (hundreds of micrometers). Conversely, TSAW-based devices do not require such precise alignment and have no inherent limit on maximum displacement. TSAW-based applications have been used to sort small particles based on differences in size, fluorescence staining, and material properties (such as density and compressibility) (Collins et al. 2016; Ma et al. 2016; Ma et al. 2017; Mutafopulos et al. 2019; Li et al. 2019; Afzal et al. 2020; Li and Ai 2021; Li et al. 2021; Nawaz et al. 2023). Our research group recently demonstrated sorting of SC-β cell clusters by size using TSAWs, therefore demonstrating for the first time label-free sorting of large biological systems by size using TSAWs (Sethia et al. 2024). This sorting of SC-β cell clusters was achieved at throughputs of up to 0.2 SC-β cell clusters/s. However, there is still a need to understand the role of various control parameters in sorting and to optimize them for high-throughput and high-efficiency sorting of larger size biological systems and particles.

In this work, a TSAW-based microfluidic platform was used to separate polyethylene particles by image-based detection. This was done at a throughput that was orders of magnitude better than previous research-based systems using image-based detection and comparable to much more costly and bulky commercial separation systems. The TSAW-based microfluidic device used in this study is specifically relevant for pancreatic islet transplantation and has been used to sort cell aggregates into small and large samples in our recent study (Sethia et al. 2024). It has been shown that smaller-sized islets (<150µm) are more effective at curing diabetes than larger-sized ones (>150µm) when normalized for cell number/mass (Lehmann et al. 2007). Under conditions in which the total allowable islet mass transplanted is limiting, we could use smaller islets to increase the likelihood of curing diabetes. Thus, in our current study, we used particles reflecting this size distribution to develop more insights into the large particle size sorting process. A binary mixture of small (45-53 µm) and large (150-180 µm) particles were used in this study for easier visualization and monitoring of the sorting process. Particles larger than the user-chosen size cutoff (>150$\mu$m, which represents larger islets) were separated using TSAW pulses. The limits of the designed actuation device were systematically explored by varying the control parameters, including flow rates and TSAW pulse power and duration. We used this systematic exploration to characterize the sorting process, and in this way, high-throughput, and high-purity sorting for a wide range of particle compositions were met.

## 2. Methodology

### 2.1 Operating principle

Understanding particle motion due to a propagating acoustic field has been an area of great interest (King 1934; Hasegawa and Yosioka 1969; Settnes and Bruus 2012). Non-linear propagation and dissipation of the acoustic field in the fluid leads to the generation of acoustic streaming flow (ASF). The acoustic radiation force (ARF) acts on the particles to push them along the direction of wave propagation in the fluid. The time-averaged ARF acting on a particle can be estimated using the equation $\langle ARF \rangle = \pi r^2 Y_p \langle E \rangle$, where $r$ is the radius of the particle, $\langle E \rangle$ is the wave's time-averaged energy density, and $Y_p$ is the acoustic radiation factor, which is dependent on the particle size and material properties (Hasegawa and Yosioka 1969). It has been found that ARF can only sufficiently translate the particles when they are larger than a specific size. ARF action on a particle is governed by Helmholtz's number ($\kappa$), which is defined as $\kappa = 2\pi r/\lambda_f$, where $\lambda_f$ is the acoustic wavelength in the fluid. For $\kappa \lesssim 1$, isotropic scattering occurs, whereas for $\kappa \gtrsim 1$, anisotropic scattering of the TSAWs leads to a net transfer of momentum to the particles (Skowronek et al. 2013; Destgeer et al. 2014; Destgeer et al. 2015; Collins et al. 2016). Since all particles used in our experiments were ≥45$\mu$m in size, the $\kappa$ value is above 1.9, implying that ARF could act to push all particles-used along the direction of propagation of the wave. Further details on this can be found in S.1-S.2.

### 2.2 Feedback-based sorting

The image recognition–based sorting of the particles is demonstrated in Figure 1. The microfluidic device is mounted over an inverted microscope (IX73, Olympus Corporation, Japan) and observed using a suitable (2x) magnification. The particle mixture is introduced into the device through inlet I2 and is laterally sandwiched using two sheath flows entering through inlets I1 and I3. The focused stream of particles then enters an optical interrogation zone where particle images are captured using a camera (acA2040-120um, Basler, Germany). The captured images are then received by a custom-written Python-based graphical user interface (GUI) to identify and track particle motion. Processes performed sequentially using the GUI are summarized in Fig. S2. The GUI signals the remotely connected waveform generator (33519B, Keysight Technologies, Santa Rosa, CA) to generate an AC pulse with the user-chosen frequency, power, and duration. The pulse is amplified using a power amplifier (LZY-22+, Mini-Circuits, Brooklyn, NY) and finally delivered to the IDTs. The piezo wafer beneath the IDTs transforms the incoming electrical signals into mechanical waves of the same frequency. These waves then travel to the bonded microfluidic device and leak into the fluid, pushing the larger particle towards outlet O2. Particles smaller than the user-chosen size cutoff do not trigger IDTs, and therefore the smaller particles are intended to continue their motion and exit the channel through outlet O1. Further details on device fabrication and sample preparation are provided in the S.3-S.4.

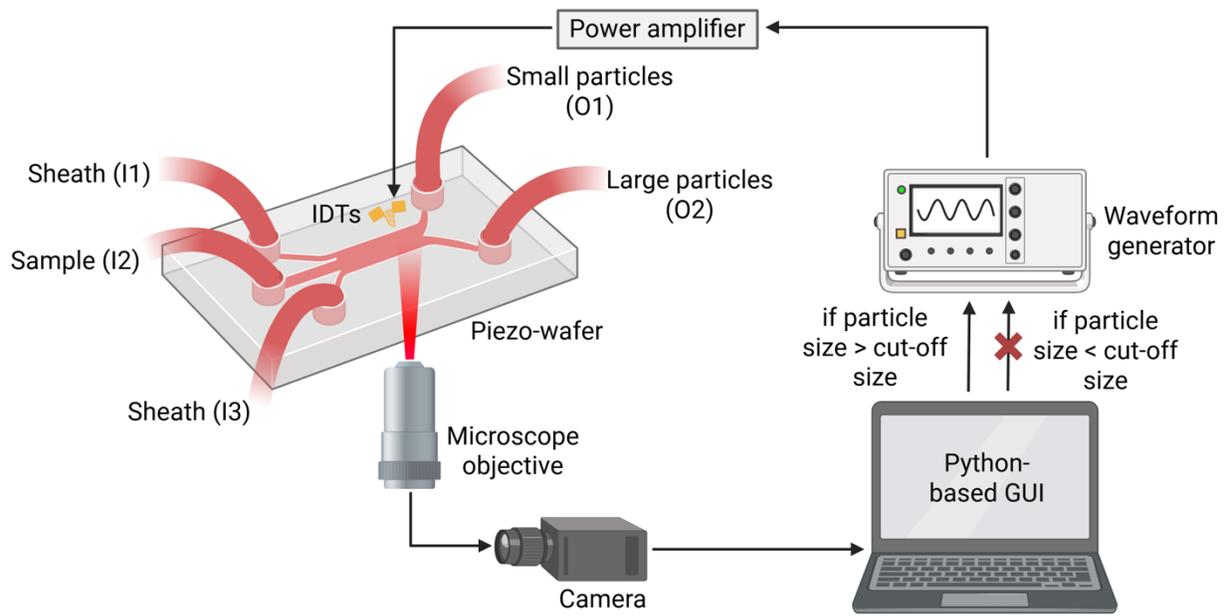

(a)

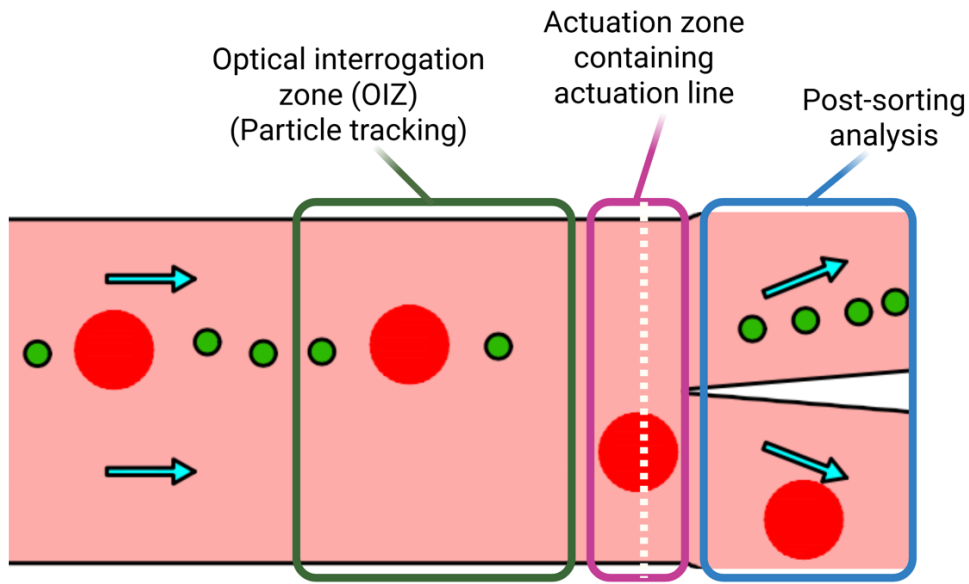

(b)

*Fig. 1* Schematic illustration of the device design and the system setup. (a) The mixture of varying particle sizes (I2) is focused using two sheath flows (I1, I3). Particles pass through an optical interrogation zone where they are recorded using a camera. The captured images are sent to a *custom-written Python-based graphical user interface (GUI), which detects the position, size, and speed of the particles. The time for the particle to reach the actuation line (under the TSAW working area) is estimated for all particles larger than the user-chosen size cutoff. Next, the GUI signals a remotely connected waveform generator to generate a wave of a given frequency, power, and duration when a particle approaches the actuation line. The wave is amplified using a power amplifier and is delivered to the interdigital transducers (IDTs). IDTs act to generate TSAWs to push the identified particles towards outlet O2. The unactuated particles continue to flow toward outlet O1. (b) Top-view images of micro-channels, close to the channel exit, are captured and processed during a typical sorting experiment. During sorting, each particle passes through various zones, labeled in the figure using colored boxes. First the particle motion is monitored in the optical interrogation zone. If the criterion for actuation is met, then it is translated laterally near the actuation line (within the actuation zone). Videos of the three regions are further analyzed offline to determine the number of particles exiting through each outlet. Green circles represent small particles, and red circles represent large particles. Figures not to scale. (Created with BioRender.com)*

**2.3 Sorting recovery evaluation**

The number of particles exiting each outlet were counted, using recorded videos of each experiment, to determine the system's sorting efficiency and total throughput. The output metric %S recovery was defined as the percentage of small particles exited through their designated outlet, O1, and %L recovery was the percentage of large particles exiting through their designated outlet O2:

$$\text{\%S recovery} = \frac{\text{Small particles exiting (O1)}}{\text{Small particles exiting (O1 + O2)}} \times 100 \qquad (1)$$

$$\text{\%L recovery} = \frac{\text{Large particles exiting (O2)}}{\text{Large particles exiting (O1 + O2)}} \times 100 \qquad (2)$$

Additional details on particle tracking and counting have also been provided in the S.5-S.6.

## 3. Results

### 3.1 Actuation of large particles using TSAW

A critical first step to sorting was to demonstrate the ability to actuate large particles with TSAWs. The effectiveness of the designed TSAW-based system to identify and actuate the particle while in flow was monitored. Large particles (150-180$\mu$m) were first focused into narrow streams with the help of sheath flows. The flow rates of the upper sheath (I1), particle flow (I2), and lower sheath (I3) were 88$\mu$L/min, 20$\mu$L/min, and 150$\mu$L/min, respectively. These flow rates allowed the particle centers to be focused ~540$\mu$m from the channel side wall closer to the IDTs, requiring ~160$\mu$m displacement to reach the bifurcation line (passing through the bifurcation point). These particles were then actuated with a single TSAW pulse of chosen duration and power in the actuation zone. Upon actuation, particles moved away from the closest side wall, toward the bifurcation line. As anticipated from previous studies (Destgeer et al. 2014; Ma et al. 2017) (of <10$\mu$m particle actuation), the particles were pushed further with increased duration and power of the TSAW pulses. This observation was novel, confirming that particles >150$\mu$m could also be increasingly translated using longer and higher-power TSAW pulses of comparable wavelength (200$\mu$m). At this wavelength, TSAW delivers energy at high frequency (~20MHz), thereby reducing the time to actuate (~1ms), which is necessary for high-throughput sorting.

The pulse duration required to push most of the larger particles over the bifurcation line was a function of the power of the TSAWs used. As shown in Fig. 2a, 0.7ms and 0.9ms pulses were sufficient to push particles to the other outlet (O2) across the bifurcation line using 36.5dBm and 39.0dBm power, respectively. Actuation of smaller particles (45-53$\mu$m) was also tested using the 36.5dBm-0.9ms pulse, as this power-duration combination was used for most of the subsequent experiments. Given particles have $\kappa$ values >1.9, actuation of particles of both sizes by ARF was expected (as discussed in Sec. 2.1), which was also experimentally confirmed. As representatively demonstrated in Fig. 2b, smaller particles undergo smaller displacements (177±35$\mu$m) compared to larger particles (217±50$\mu$m). This difference can be attributed to the different particle acoustic radiation factors ($Y_p$) and sizes ($r$) of small and large particles. ARF acting on particles increases proportional to $r^2 Y_p$, where $Y_p$ is a nonlinear function of $r$ (as detailed in Fig. S1).

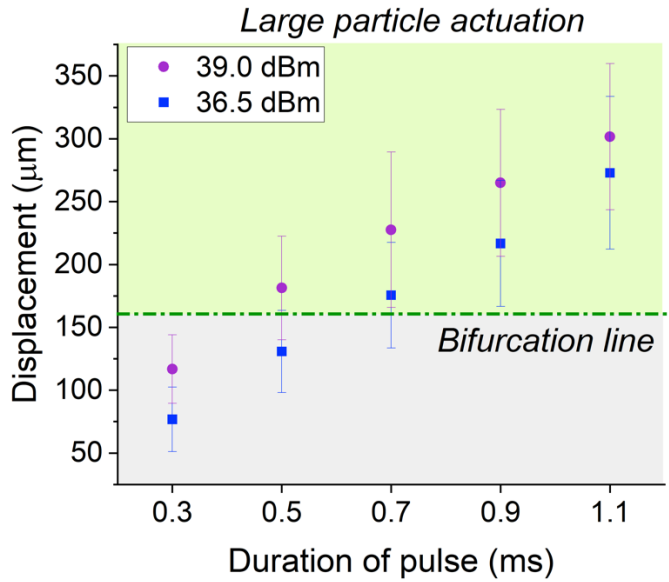

(a)

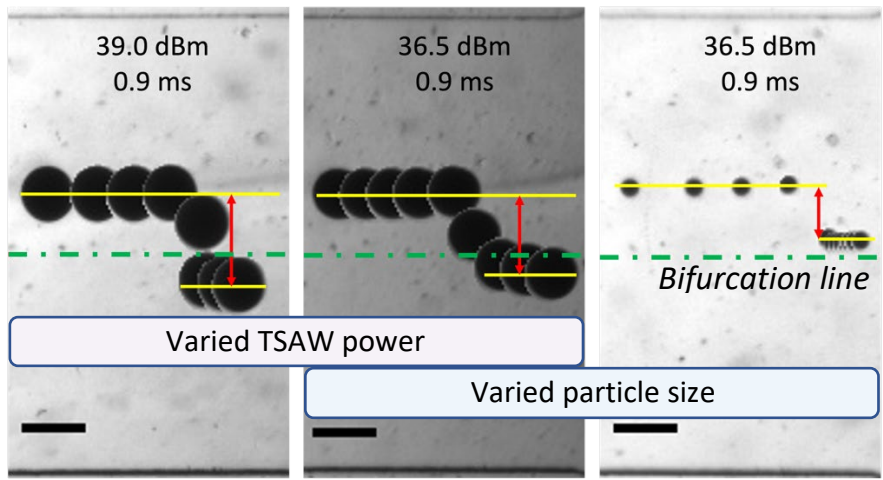

(b)

***Fig. 2*** *Actuation of particles under varying duration and power of TSAW pulse. (a) Focused larger particles (150–180μm) are actuated using TSAW pulses of varying powers (36.5 dBm and 39.0 dBm) and duration (0.3–1.1ms). Displacement of particles, defined as the center-to-center distance moved by the particle transverse to the main flow direction, was measured for each pulse power-duration combination with the help of a custom-written Python code. Average particle displacement increases with both the power and the duration of the TSAW pulse acting on it. The*

*dotted green line (denoting crossing over the bifurcation line) marks the distance by which the particle needs to be displaced on average for it to exit through the outlet O2. Error bars represent standard deviation in the displacement of particles for each pulse power-duration combination (n>50 per combination). (b) Time-lapsed particle positions (overlapped over the first frame) showing how the displacement is measured for (a). Smaller particles also actuated using TSAW pulse (0.9ms, 36.5 dBm) show smaller displacement compared to the larger particles. Raw images were processed to generate time lapsed images using a custom-written Python code. Scale bar is 200μm*

**3.2 Effect of actuation signal–related parameters on sorting**

The sorting of a prepared mixture of small and large particles was tested using different durations of TSAW pulses at the chosen power (36.5dBm). As shown in Fig. 3, for 0.5ms and 0.7ms pulse duration, there was a low %L recovery, implying large number of the large particles were unable to cross the bifurcation line to enter the outlet O2. For pulse durations ≥0.9ms, more than 97% of large particles were sent to the correct outlet, which was expected based on Fig 2a. A slight decrease in %S recovery with the increased pulse duration was also observed, implying increasingly small particles exited through outlet O2 instead of outlet O1. This was because the small particles that were close to large particles were also acted upon by the ARF, so they were pushed alongside large particles. Using the results from Fig 3, 0.9ms was chosen as the desired pulse duration for the sorting. At this pulse-power combination, large particles will be sufficiently pushed off while maintaining high recovery (~97%) of small particles from outlet O1. The pre-sorted mixture had ~0.5% w/v concentration of each size particle. By number, large particles were ~2-5% of the total particles, matching the fraction of large islets (>150$\mu$m) in a typical islet isolation (Korsgren et al. 2005). Inlet flow rates were identical to the implementation in Fig. 2a (I1: 88$\mu$L/min, I2: 20$\mu$L/min, and I3: 150$\mu$L/min). Additionally, IDTs were able to actuate particle mixtures (using this pulse-power combination) across the width of the actuation zone (more details in S.7).

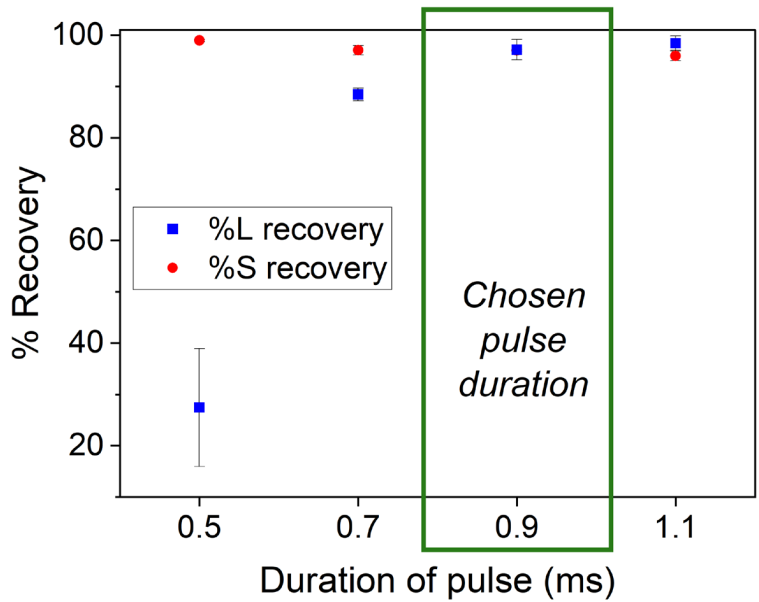

*Fig. 3 Separation of particle mixtures under varying TSAW pulse durations. For the chosen power (36.5 dBm), duration of TSAW pulse was varied from 0.5 to 1.1ms to actuate off the large particles (>150μm) from the mixture. As expected, a 0.9ms pulse was able to successfully push (97$\pm$2%) large particles to their designated outlet, O2. Most (97$\pm$0%) of the small particles in the mixture successfully exited the channel through their designated outlet, O1. Inlet flow rates were kept constant at 88μL/min for I1, 20μL/min for I2, and 150μL/min for I3. Error bars represent standard*

*deviation in % recoveries of particles as calculated from n=3 experimental videos per pulse duration*

**3.3 Effect of increasing flow rates on sorting of large particles**

Flow rates were increased, with each inlet flow rate scaled by the same factor relative to 1x flow rates (I1: 88$\mu$L/min, I2: 20$\mu$L/min and I3: 150$\mu$L/min), to further increase the throughput of the device. Channel Reynold's number, defined as $Re = \frac{\rho_f V_{avg} H}{\mu_f}$, where $\rho_f$, $\mu_f$, $V_{avg}$, and $H$ are the fluid density, fluid viscosity, average channel cross-sectional velocity, and channel height, respectively, varied in the range ~ 3.1–10.8. This range implied laminar flow. As the flow rate increased, % recoveries of particles varied due to multiple competing events in the system. Notably, particle focusing using sheath flows, latency in particle actuation, and particle-trajectory post-actuation affect sorting recoveries at higher flow rates. As shown in Fig. 4a, large-particle recovery is lowest when 2.5x flow rates are used in the system. Recoveries on either side of 2.5x flow rates improve due to fewer latency challenges (at lower flow rates) and lateral focusing of particles closer to the bifurcation (at higher flow rates). On the other hand, smaller particle recoveries do not benefit much from increased flow rates. As shown in Fig. 4b, more small particles are inadvertently pushed into outlet O2 with increasing flow rates.

The small-particle recovery remains at ~97% up to 3x inlet flow rates, beyond which it decreases significantly. As observed previously, small particles can also be present in proximity to large particles within the actuation zone. At lower flow rates, such small particles might be actuated but not necessarily traverse to outlet O2, due to their shorter displacements as shown in Fig. 2b.

Post-actuation, the satellite small particles slow down due to being pushed closer to the top channel wall as reported previously (Destgeer et al. 2014; Collins et al. 2016) (and summarized in S.2). At higher flow rates, due to more frequent actuations, such slowed-down particles can be actuated multiple times. The multi-actuated small particles can then traverse distances sufficient to exit through outlet O2. It was also observed that at higher flow rates, particles focused into the streamlines laterally closer to the bifurcation (Fig. S4). This also made it easier for the satellite small particles to enter outlet O2, due to reduced displacement requirements.

Increasing flow rates leave a smaller time window for the tracking and actuation of particles, and therefore higher image acquisitions rates were used. At higher flow rates, some signals were late due to inherent lag in the system (signal delivery from GUI to waveform generator). This delayed signal delivery resulted in large particles exiting through outlet O1 (instead of designated outlet O2). Approximately 42% and 85% of the cumulated number of large particles that exited outlet O1 at 1x and 3.5x flow rates were found not to be actuated in a timely manner respectively. However, the actuation line was varied within the actuation zone to accommodate for the higher flowrates used, and therefore the increase in the number of particles not actuated in timely manner were not due to the positioning of the actuation line. On the other hand, as particles focused laterally closer to the bifurcation line at higher flow rates, they were easier to actuate. Due to these competing events, large-particle recovery is the lowest at 2.5x flow rates, beyond which particle focusing compensates for the delayed actuation signal. Across all inlets, a 3x inlet flow rate (I1: 264$\mu$L/min, I2: 60$\mu$L/min and I3: 450$\mu$L/min) was chosen for further experimentation.

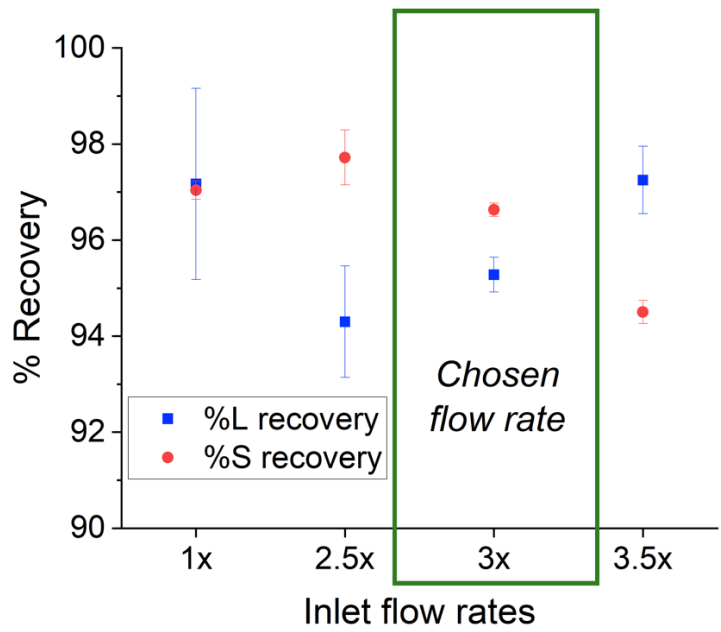

**(a)**

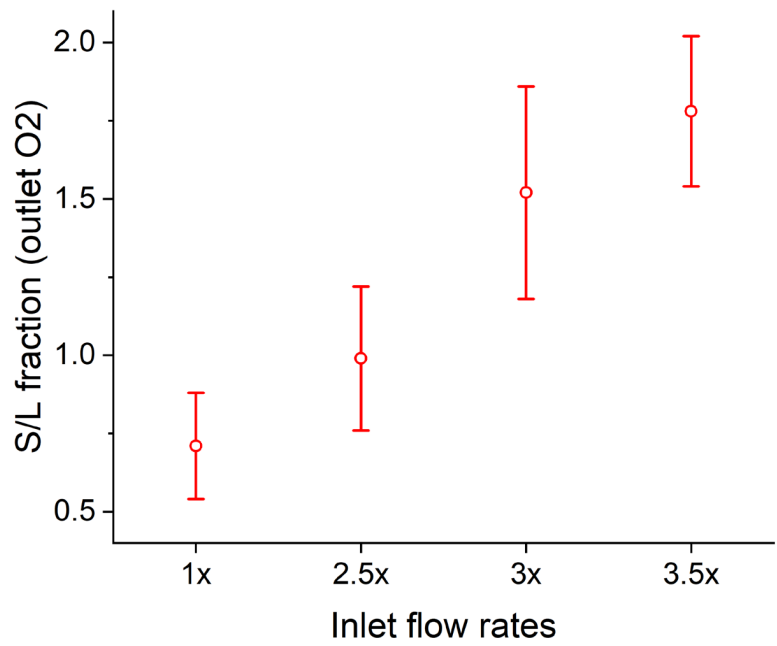

**(b)**

*Fig. 4* *Increasing sorting throughput using higher flow rates. (a) Each inlet flow rate was scaled up while keeping the ratio between them constant. All inlet flow rates were increased by multiples of 2.5–3.5 relative to the previously used flow rates (for data shown in Figures 2 and 3) of I1: 88µL/min, I2: 20µL/min and I3: 150µL/min (labeled as 1x). On average, particle recoveries remain above 94%, even though there is a trade-off between %L and %S recoveries. (b) With increasing flow rates, more small particles are also inadvertently pushed into outlet O2 (meant for large-particle collection). Consequently, the small-to-large particle ratio in outlet O2 also increases with flow rate. Particles were actuated using the chosen (36.5 dBm/0.9ms) TSAW pulse. Error bars represent standard deviation in % recoveries of particle as calculated from n=3-4 experimental videos per flow rate*

### 3.4 Sorting wide compositions of pre-sorted particle mixtures

Thus far, the pre-sorted particle mixtures had a large-particle concentration of ~0.5% w/v, as they constitute ~2–5% of all particles by number. This ratio was chosen to reflect the expected fraction of large islets (>150$\mu$m) in a typical human pancreatic islet isolation (Korsgren et al. 2005). To further accommodate for possible biological variations and design a more generalized sorting technique, sorting experiments with increasing large-particle fractions were performed. First the concentration of large particles in the pre-sorted particle mixture was quadrupled (up to 2.0% w/v) while keeping small particle concentration to 0.25% w/v. On one hand, at higher particle concentrations, one can expect an increasing throughput and more frequent actuations. On the other hand, particle recoveries can decrease due to the latency in the system limiting actuating frequency or due to increasing particle proximities and overlaps. At the chosen large-particle concentration of ~1.0% w/v, particle mixtures with large particles making up to ~8–45% of the total particles by number on average were sorted. Interestingly, the %recovery of large and small particles remained consistent for a broad range of particle compositions.

The large particle concentration (% w/v) in the pre-sorted particle mixture was first increased, and its effect on particle recoveries was observed. The small-particle concentration was kept constant. As shown in Fig. 5a, the small-particle recovery decreased with increased large-particle concentration. As the samples became more dominated by large particles, more actuations took place, and correspondingly, a higher number of smaller particles were removed. Large-particle recovery remained constant (>95%) till ~1.0% w/v large particle concentrations were used. The recovery rate decreased on further scale-up to ~2.0% w/v concentration. This decrease could be related to difficulty in precisely actuating closely spaced particles at higher particle concentrations. At all the concentrations studied, the final small-to-large particle (S/L) fraction in the outlet O2 stream remained nearly constant at ~0.7. Actuation frequency was estimated as total number of large particles flown through device per unit time. Overall, average actuation frequency was ~1.5 actuations/s at ~0.5% w/v concentration, ~3 actuations/s at ~1.0% w/v concentration and ~6 actuations/s at ~2.0% w/v concentration respectively.

Since both small and large particle recovery remained over 93% up to 1.0% w/v large-particle concentration, experiments on a wide range of pre-sorted compositions at constant 1.0% w/v large-particle concentrations were carried out. Average small- and large-particle recovery did not vary greatly, remaining at ~93–94% and ~94–97%, respectively, for various pre-sorted compositions (Fig. 5b). Movie S1 representatively demonstrates how small and large particles were separated into their respective outlet at one such composition. Small-particle recovery depends on two counteracting factors. At higher large-particle compositions, there are fewer small particles in flow and therefore fewer small particles in the vicinity of large particles. On one hand, an increase in small-particle recovery can be expected as fewer small satellite particles might actuate alongside larger particles (S/L fraction in outlet O2 decreases). On the other hand, fewer smaller particles exited outlet O1 between successive large-particle actuations, counteracting the effect of fewer satellite particles and resulting in a decrease in small-particle recovery.

Hence, it was demonstrated that our TSAW-based sorting system could process a wide range of particle mixtures (large particles constituting on average ~8–45% of the total particles), without a substantial difference in sorting recoveries. Overall, the average sorting rate varied with composition and was in the range of ~4.9–34.3 particles/s (with an average of ~2–3 actuations of large particles occurring every second). These results suggest that, depending on the required recoveries and throughput, one can dilute the pre-sorting sample in the range of 0.5–2% w/v large particle concentration and expect a %L recovery of ~92–97% and %S recovery of ~87–96%, regardless of the initial sample composition (for <0.25% w/v small particle concentration).

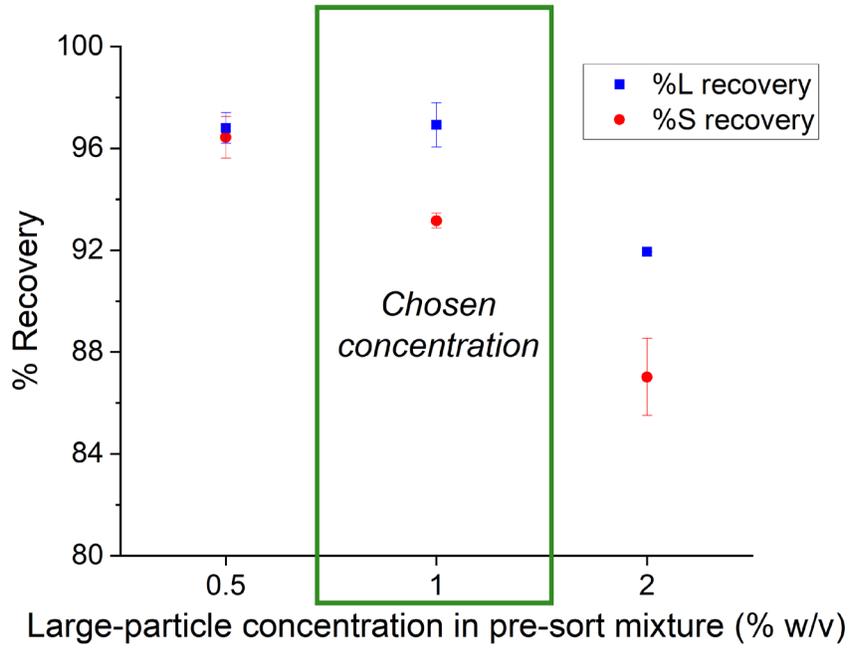

(a)

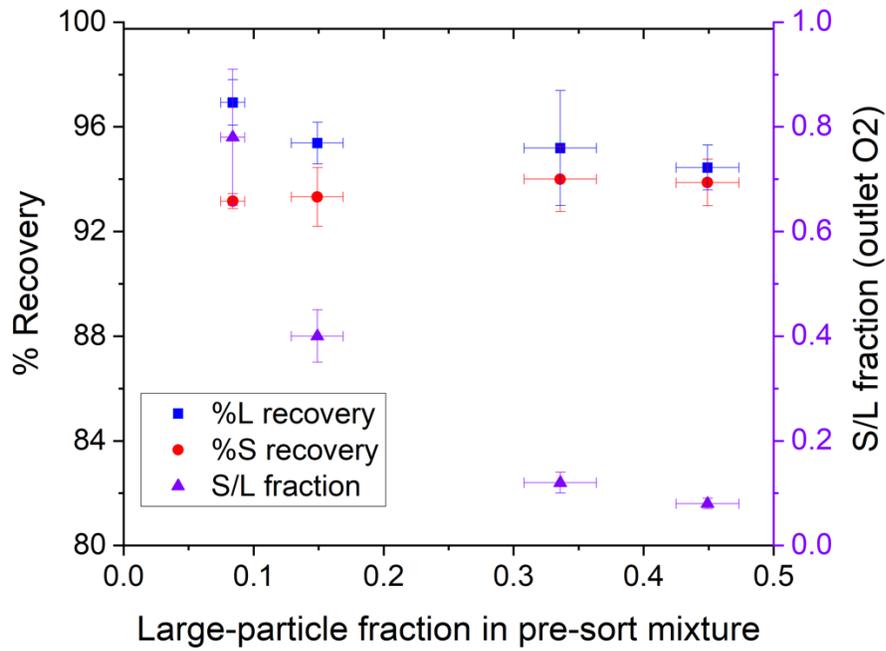

(b)

***Fig. 5*** *Ability to sort varying concentration and composition particle mixtures. (a) Experiments were performed with increasing large-particle concentrations (0.5–2.0% w/v). More-concentrated solutions require frequent actuation but can also increase the total throughput of the system. Small and large particle recoveries remain over 90% when the concentration is increased up to 1.0% w/v. Small-particle concentration was kept constant (0.25% w/v). At all concentrations studied, the final small-to-large particle fraction in the outlet O2 stream remained relatively constant at ~0.7. (b) Particle mixtures of varying composition (up to 45% large particles on average in the pre-sorted mixture) were sorted with over 93% small- and large-particle recoveries. Large-particle concentration was kept constant (1.0% w/v). Chosen 3x inlet flow rates were used (I1: 264µL/min, I2: 60µL/min and I3: 450µL/min), and particles were actuated using the chosen (36.5dBm/0.9ms) TSAW pulses. Error bars represent standard deviations in % recoveries of particles, small-to-large particle fraction in the outlet O2 and particle mixture composition as calculated from n=3-4 experimental videos per concentration and per composition of particle mixtures*

## 4. Discussion

Currently, the average actuation frequency at the chosen conditions (36.5dBm-0.9ms TSAW pulse/3x flow rates/1.0%w/v large-particle concentration) was ~2–3 actuations/s, with ~4.9–34.3 particles sorted/s. A throughput of ~12 particles/s can sort ~ 1 million pancreatic islets within a day. Here, the efficiency of sorting may be further improved by designing IDTs with a smaller working area. Assuming uniform actuation of small particles across the working area (as seen in Fig. S3), a 2-fold decrease in working area (to ~380$\mu$m) can possibly increase the small particle recovery to ~97%. This could possibly be done by curving the IDTs further. Actuation regions as small as ~25$\mu$m wide have been designed by curving the IDTs so that they subtended an angle of 26° at the center (Collins et al. 2016). These focused IDTs can then be equipped to actuate each particle precisely with TSAW pulses of much lower power (about 50% decrease as wave amplitude $\propto \sqrt{\text{(power/working area)}}$) (Schmid et al. 2012). Throughput of the device could then be further increased up to theoretical maximum of ~1000 actuations/s, based on sub-millisecond actuation pulse. To further increase throughput, multiple devices can be run in parallel, or to further increase the particle recovery, multiple passes of the sample can be done till desired purity is achieved. This study can be used to improve SC-β cell clusters sorting efficiency and throughput achieved in our previous study (Sethia et al. 2024). Finally, this technique of sorting particles can easily be translated to sorting intact pancreatic islets by size, which could potentially improve islet cryopreservation and transplant outcomes (Suszynski et al. 2014; Zhan et al. 2022).

## 5. Conclusion

This work demonstrates a high-throughput sorting technique for label-free detection over a piezo-based actuator. Sorting a wide range of large-particle (45–180$\mu$m) compositions into two size groups was showcased. The work is relevant for sorting microplastics and biomaterials such as pancreatic islets. We explored the effect that control parameters, including actuation signal, flow rates, and particle concentration, have on the success of sorting. In doing so, the underlying mechanisms governing particle separation and associated purity–throughput trade-offs are elucidated. First, particles were acted upon by a high-frequency (~20MHz) TSAW pulse. Since particles had $\kappa$ values above 1.9, ARF was the dominant force that pushed the particles along the direction of wave propagation. A sub-millisecond (0.9ms) TSAW pulse that led to more than 97% particle recovery was identified. Then, throughputs were increased by threefold by scaling up of the inlet flow rates. Lastly, a broad range of particle concentrations and compositions were sorted to demonstrate the versatile nature of the device. The device works robustly and can separate any sample volume. This work is a first step towards optimizing sorting protocol for high efficiency and high throughput sorting of biomaterials, such as pancreatic islet and SC-β cell clusters, by size using this device.

**Acknowledgements**

We acknowledge and thank Sanjeeth Tirukavalluri for helping in data analysis in Fig. 3-5 and Fig. S3. We also acknowledge and thank Meenal Rathi for helping with the literature review. This work is supported by grants from the National Institutes of Health (R01DK131209, EBF, JCB) and the National Science Foundation (EEC 1941543, JCB, EBF). The device fabrication work was supported by National Science Foundation under Award Number ECCS-2025124.

**Data availability**

Data used and/or analyzed in the paper can be made available from the corresponding author on a reasonable request.

**Statements and Declarations**

The authors declare no competing interests.

**Supplementary information**

Supplementary information for this article is enclosed.